\documentclass[aip,nofootinbib,reprint]{revtex4-2}

\usepackage{amsmath}
\usepackage{amssymb}
\usepackage{empheq}
\usepackage{mathrsfs}
\usepackage{xcolor}				
\usepackage{amssymb}

\newcommand{\Hc}{\mathcal{H}_{\rm{c}}}
\newcommand{\Hp}{\mathcal{H}_{\rm{p}}}
\newcommand{\tHp}{\tilde{\mathcal{H}}_{\rm{p}}}
\newcommand{\bHp}{\check{\mathcal{H}}_{\rm{p}}}
\newcommand{\barHp}{\bar{\mathcal{H}}_{\rm{p}}}
\newcommand{\hHp}{\hat{\mathcal{H}}_{\rm{p}}}

\newcommand{\pr}{p_r}
\newcommand{\prho}{p_{\rho}}
\newcommand{\pu}{p_u}
\newcommand{\Ei}{\mathrm{Ei}}

\begin{document}


\title{Chaotic Dynamics Driven by Particle-Core Interactions}

\author{Konstantin Batygin}
\email{kbatygin@caltech.edu; author to whom correspondence should be addressed.}
\affiliation{Division of Geological and Planetary Sciences California Institute of Technology, Pasadena, CA 91125, USA}

\author{Yuri K. Batygin}
\email{batygin@lanl.gov}
\affiliation{Los Alamos National Laboratory, Los Alamos, NM 87545, USA}

\begin{abstract}
High-intensity beams in modern linacs are frequently encircled by diffuse halos, which drive sustained particle losses and result in gradual degradation of accelerating structures. In large part, the growth of halos is facilitated by internal space-charge forces within the beams, and detailed characterization of this process constitutes an active area of ongoing research. A partial understanding of dynamics that ensue within space-charge dominated beams is presented by the particle-core interaction paradigm -- a mathematical model wherein single particle dynamics, subject to the collective potential of the core, are treated as a proxy for the broader behavior of the beam. In this work, we investigate the conditions for the onset of large-scale chaos within the framework of this model, and demonstrate that the propensity towards stochastic evolution is strongly dependent upon the charge distribution of the beam. In particular, we show that while particle motion within a uniformly charged beam is dominantly regular, rapid deterministic chaos readily arises within space-charge dominated Gaussian beams. Importantly, we find that for sufficiently high values of the beam's space charge and beam pulsation amplitude, enhanced chaotic mixing between the core and the halo can lead to an enhanced radial diffusion of charged particles. We explain our results from analytic grounds by demonstrating that chaotic motion is driven by the intersection of two principal resonances of the system, and derive the relevant overlap conditions. Additionally, our analysis illuminates a close connection between the mathematical formulation of the particle-core interaction model and the Andoyer family of integrable Hamiltonians.
\end{abstract}

                              
\maketitle

\section{Introduction} \label{sec:intro}
An important and long-standing challenge to the performance of high-intensity particle accelerators is presented by the emergence of halos around space-charge dominated beams. Simply put, a beam halo is comprised of a tenuous population of charged particles that fill a phase-space volume that is more extended than that occupied by the beam's core. Within a halo, particles can execute long-range radial excursions, leading to their eventual absorption by the accelerating structure. Correspondingly, this process limits the intensity of the beam and results in radio-activation of the linac\cite{1998rla..book.....W}.

Although the aforementioned practical issues are broadly appreciated \cite{Jameson1993,W98,Qiang2004}, efforts to suppress halo formation through beam collimation cannot resolve the problem completely, as halos have a distinct tendency to regenerate over relatively short timescales. Fundamentally, this propensity towards reemergence indicates that intrinsic self-excitation of the beam through non-linear space-charge forces plays an appreciable, if not dominant, role in the process of halo formation \cite{G94,L94a}. In this vein, one potential suppression mechanism for beam halo formation employs nonlinear focusing, as originally proposed in \cite{1996PhRvE..53.5358B,1998PhRvE..57.6020B} and further developed in \cite{2005PhRvS...8f4202S,2017PhRvS..20a4201P}. Nevertheless, a detailed understanding of this halo-formation mechanism remains incomplete, and constitutes an active area of investigation \cite{L94b,2001pac..conf.1732A,Gerigk2004,2007NIMPA.577..173L}.

One illustrative approach towards characterizing phase-space growth of space-charge-dominated beams rests on the \textit{particle-core interaction model} of halo formation \cite{G94,W98}. Within the context of this picture, the evolution of a single test-particle -- subject to the collective potential of a breathing axisymmetric beam -- is treated as a proxy for typical dynamics exhibited by charged particles. More specifically, the radial oscillations of the beam are taken to stem from an integrable Hamiltonian, while the motion of the test-particle is governed by a non-autonomous Hamiltonian with only a single degree of freedom \cite{Ikegami1999}. For typical parameters, this model yields a pulsating core that is encircled by a separatrix, which can transport particles from the outer edge of the core to a considerably larger radial distance, thereby creating a diffuse halo.

Despite the particle-core interaction model's apparent lack of complexity, it has been shown to readily yield non-trivial  behavior \cite{W98,BV}, and has displayed a noteworthy degree of agreement with experimental data \cite{2002PhRvL..89u4802A}. Moreover, the model's capacity to capture the transition from ordered, quasi-periodic motion to deterministic chaos renders it an ideal paradigm for exploring halo formation through stochastic transport of particles. We note, however, that even though the appearance of nonlinear resonances and the materialization of chaotic bands in phase-space have been routinely observed in numerical simulations of the particle-core interaction model, the phase-space attributes of these resonances, the specific conditions for the onset of large-scale chaos, as well as  the dependence of the beam's stochastic behavior on its radial charge distribution remain to be quantified from analytic grounds. 

Carrying out this analysis is the primary purpose of this paper. More specifically, a key aim of this paper is to employ canonical perturbation theory to systematically explore two (uniform and Gaussian) variants of the particle-core interaction model, with an eye towards identifying the principal resonant dynamics that facilitate halo formation. The remainder of the manuscript is organized as follows. We begin our analysis in section \ref{sec:unichargedbeam} by reviewing particle motion within a uniformly-charged beam. We then explicitly identify the primary harmonic, responsible for large-amplitude radial excursions, through time-averaging of the Hamiltonian. With a simplified model in hand, we highlight a hitherto-overlooked connection between the particle-core interaction model and the so-called ``second fundamental model for resonance" used extensively in celestial mechanics \cite{HL83}. Subsequently, we derive the associated resonance proximity parameter in terms of the beam's current and pulsation amplitude. 

In section \ref{sec:gaussianbeam}, we consider the particle-core interaction model with a Gaussian charge distribution, and repeat our perturbative analysis, this time carrying out the time-averaging procedure semi-analytically. By comparing our results with stroboscopic surfaces of section, we demonstrate that the beam core of a Gaussian variant of the model is markedly more susceptible to chaotic mixing, and outline a criterion for the onset of chaos through an analogy with a modulated pendulum. We further demonstrate that resonant excursion of particles to large radii can be sourced from deep within the core. Finally, we confirm our results with a suite of numerical simulations, and identify a parameter range where halo formation is strongly enhanced by long-range stochastic diffusion. We briefly summarize our findings in section \ref{sec:summary}. 

\section{A Uniformly Charged Beam} \label{sec:unichargedbeam}

The mathematical formulation of the particle-core interaction model amounts to two second order (equivalently, two pairs of first order), coupled non-linear ODEs: one for the radial pulsation of the beam core and the second for the radial motion of the test-particle. As already alluded to in the previous section, however, the coupling between these equations is uni-directional, such that oscillations of the core around its equilibrium radius, $R$, modulate the particle's phase-space trajectory, but not vice-versa. Accordingly, let us begin by quantifying the core's radial breathing as a function of time.

\subsection{Pulsation of the Core} \label{sec:pulsecore}


The dynamics of the beam core's radius is governed by the Hamiltonian \cite{G94}:
\begin{align}
\Hc=\frac{\pr^2}{2}+\frac{r^2}{2}+\frac{1}{2\,(1+b)}\frac{1}{r^2}-\frac{b\,\ln(r)}{1+b},
\label{eqn:Hcore}
\end{align}
where $r=x_{\rm{c}}/R$ is a dimensionless core radius and $\pr=\dot{r}$ is the conjugate momentum. Strictly speaking, this mono-dimensional description of the beam's evolution entails axial symmetry, and is valid for an axially-symmetric beam in a continuous solenoidal focusing channel. In a quadrupole channel, on the other hand, equation (\ref{eqn:Hcore}) describes the evolution of a smoothed envelope, averaged in both $\hat{x}$- and $\hat{y}$-directions, wherein small-amplitude fast oscillations are filtered out. We note that although the inclusion of 2D or 3D pulsation (see e.g., \cite{2000PhRvS...3f4201Q}) can result in additional higher-order dynamics, such effects are by no means central to the problem at hand.

The Hamiltonian is parameterized the dimensionless space-charge \cite{BV}
\begin{align}
b=\frac{2\,m\,c}{p}\frac{I}{I_{\rm{c}}}\frac{R^2}{\epsilon^2}
\end{align}
where $m$ is the particle mass, $c$ is the speed of light, $p$ is the particle momentum, $I$ is the beam current, $I_{\rm{c}} = 4\pi\,\epsilon_0\,m\,c^3/q$ is the characteristic beam current, $\epsilon_0$ is the permittivity of free space, and $\epsilon$ is the normalized beam emittance. Note that instead of $b$, \citet{W98} parameterize $\Hc$ in terms of the in-core space-charge tune-depression ratio 
\begin{align}
\frac{\sigma}{\sigma_0}=\frac{1}{\sqrt{1+b}},
\end{align}
where $\sigma$ and $\sigma_0$ are the phase advances of transverse particle oscillations per focusing period, with and without space-charge forces, respectively. Moreover, notice that because $\Hc$ is expressed in terms of dimensionless variables, the unit of time must also be appropriately non-dimensional, and here, $t=\tau\,\Omega$, where $\Omega$ is the frequency of particle oscillation in a uniform focusing channel in absence of space-charge, and $\tau$ is the dimensional time.

The Hamiltonian (\ref{eqn:Hcore}) possesses a global minimum -- which corresponds to an elliptic equilibrium -- at $(r,\pr)=(1,0)$. Since we are primarily interested in variations of ($r,\pr)$ around this fixed point, it is sensible to change variables to $\rho=r-1, \prho=\pr$ (addition or subtraction of a constant to either the coordinate or the momentum always corresponds to a canonical transformation \cite{2002clme.book.....G}), and expand the Hamiltonian around the origin. Carrying out the power-series expansion to cubic order in $\rho$, we have:
\begin{align}
\Hc &\approx\frac{\prho^2}{2}+\frac{2\,(2+b)}{1+b}\,\frac{\rho^2}{2}-\frac{(6+b)}{1+b}\,\frac{\rho^3}{3} \nonumber \\
&= \frac{\prho^2}{2}+\omega^2\,\frac{\rho^2}{2}-\zeta^2\,\frac{\rho^3}{3},
\label{eqn:Hcorepower}
\end{align}
where the characteristic frequency of core oscillation is $\omega=\sqrt{2\,(2+b)/(1+b)}$, $\zeta=\sqrt{(6+b)/(1+b)}$, and we have dropped a dynamically irrelevant constant term from the expression. 

\begin{figure*}[t]
\centering
\includegraphics[width=\textwidth]{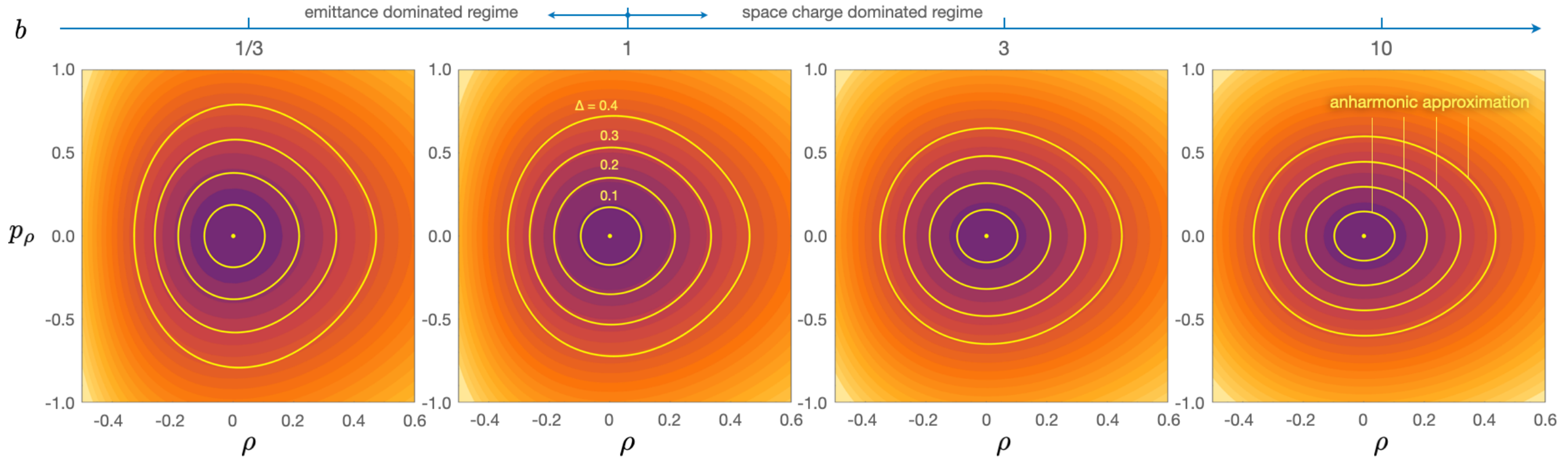}
\caption{Anharmonic approximation to pulsation of the core. The background contour plot corresponds to the level contours of Hamiltonian (\ref{eqn:Hcore}). Yellow curves are given by the approximate anharmoinc solution (\ref{eqn:anharmonic}), for oscillation amplitudes of $\Delta=0.1,0.2,0.3,$ and $0.4$. In the considered range of parameter space, equation (\ref{eqn:anharmonic}) provides an adequate approximation for the phase-space portrait of Hamiltonian (\ref{eqn:Hcore}).}
\label{F:an}
\end{figure*} 

Hamiltonian (\ref{eqn:Hcorepower}) corresponds to an anharmonic oscillator. Although a closed-form solution to the resulting equations of motion can be expressed in terms of elliptic integrals \cite{1991PhRvA..44.3484D}, for our purposes, an approximate solution for $\rho$ expressed as a rudimentary function of $t$ will be sufficient. Such a solution is readily obtained from classical perturbation theory \cite{LL,LightL} and has the form (see appendix for details):
\begin{align}
\rho&\approx \Delta \cos(\omega\, t+\varphi)+\frac{\Delta^2\,\zeta^2}{2\,\omega^2}\nonumber \\
&-\frac{\Delta^2\,\zeta^2}{6\,\omega^2}\,\cos(2\,(\omega\,t+\varphi)),
\label{eqn:anharmonic}
\end{align}
where $\Delta$ is the amplitude of oscillation, and $\varphi$ is an arbitrary phase. For consistency with previous studies \cite{L94b,W98,BV}, we set $\varphi=\pi$ such that $r$ is minimal and $\pr$ is null at $t=0$.

The solution (\ref{eqn:anharmonic}) is shown on a $(\rho,\prho)$ plane in Figure (\ref{F:an}) for a series of values of $\Delta\in(0,0.4)$ and $b\in(0.1,10)$ -- a physically relevant parameter range that we will adopt for the remainder of the paper. We note that these bounds on $\Delta$ and $b$ are motivated by experiment: even the most high-intensity machines in operation are characterized by $b\lesssim10$, while beams with oscillation amplitude in excess of $\Delta\gtrsim0.4$ typically suffer immense losses, rendering them uninteresting from a practical standpoint. On the same figure, level curves of Hamiltonian (\ref{eqn:Hcore}) are shown as the background contour map. The agreement between the approximate series solution (\ref{eqn:anharmonic}) and the evolution entailed by $\Hc$ is satisfactory for $\Delta \lesssim 0.4$ and improves for higher values of $b$. In fact, this dependence can be expected, since $(\zeta/\omega)^2$ ranges from 3/2 for $b\rightarrow0$ to $1/2$ for $b\gg1$, implying that higher order terms are slightly suppressed for space-charge dominated beams. Accordingly, for the remainder of the manuscript, we will retain equation (\ref{eqn:anharmonic}) as an acceptable approximation, but note that while the physical meaning of $r$ denotes the outer edge of the beam in case of a uniformly charged core, it corresponds to twice the rms radius of the beam in the case of a Gaussian charge distribution. 

\subsection{Particle Dynamics} \label{sec:partdyn}

With the time-dependence of $r$ specified, let us now review the dynamics of an embedded particle from a perturbative perspective. In essence, the goals of the following calculation are two-fold. First, we aim to outline the relevant approximation scheme that we will revisit in the next section. Second, by developing an approximate model, we seek to contextualize the phase-space portrait of the particle-core interaction model from purely analytic grounds. 

\subsubsection{An Integrable Approximation} \label{sec:intapp}
The motion of a particle subject to the space-charge forces of a uniformly charged beam core is governed by the following Hamiltonian \cite{G94,W98,BV}:
\begin{align}
&\Hp=\frac{\pu^2}{2}+\frac{u^2}{2}-  \frac{b}{1+b} \, \times \nonumber \\
& \begin{dcases*} 
\frac{u^2}{2\,r^2}& $|u|<r$ \\ 
\bigg(\frac{1}{2}+\ln(|u|(1+b))-\ln(r(1+b)) \bigg) & $|u|\geqslant r,$
  \end{dcases*} 
\label{eqn:Hpart}
\end{align}
where $u$ and $\pu$ refer to the particle's position and momentum, respectively. Because $\Hp$ exhibits \textit{explicit} dependence on time through $r$, it is not integrable, and analytic solutions for $(u,\pu)$ cannot exist. Conversely, numerical experiments of dynamics jointly facilitated by Hamiltonians (\ref{eqn:Hcore}) and (\ref{eqn:Hpart}) are plentiful in the literature \cite{Jameson1993,L94a,L94b,W98,Ikegami1999}, and need not be rehashed at this stage. As a result, let us instead derive an integrable approximation to Hamiltonian (\ref{eqn:Hpart}), leaving simulations for later. In spirit, our analysis will follow the same general theme as the work of \citet{G94} and \citet{BV}, although our approach to perturbation theory will be different, and will illuminate the criteria under which the averaging process is warranted.

As a starting point, we follow \citet{G94}, and replace the piecewise-continuous function in equation (\ref{eqn:Hpart}) with a smooth quartic polynomial:
\begin{align}
\Hp\approx\frac{\pu^2}{2}+\frac{u^2}{2}-\frac{b}{1+b}\bigg(\frac{u^2}{2\,r^2}-\frac{u^4}{16} \bigg).
\label{eqn:Hpartan}
\end{align}
Before proceeding further, we caution that although this polynomial fit has been suitably employed within the literature, it constitutes a relatively crude approximation for the potential function outlined in equation (\ref{eqn:Hpart}). In a purely quantitative sense, the fractional error entailed by equation (\ref{eqn:Hpartan}) is on the order of $10-20\%$ for $\Delta=0$, and can even exceed unity for $\Delta=0.4$ and $|u|\gtrsim1+\Delta$. From a qualitative perspective, it is further important to note that a strictly linear field interior to $|u| < r$ in equation (\ref{eqn:Hpart}) is replaced by a weakly non-linear one in expression (\ref{eqn:Hpartan}). Despite these shortcoming, estimates obtained within the context of this analytic model provide a satisfactory match to the results of numerical experiments (e.g., the radial extent of the resonant separatrix is reproduced to within $\sim10\%$, etc; \cite{BV}). Nevertheless, these approximations are sufficiently pronounced for us to emphasize that any results derived from Hamiltonian (\ref{eqn:Hpartan}) -- including the reduction of the problem to the Andoyer model outlined in section \ref{sec:andoyer} below -- should not be mistaken for a quantitatively accurate description of particle dynamics. Rather, the key goal of this analysis is to provide an intelligible ``blueprint" for particle dynamics, within the context of which the topological character of the phase-space portrait can be interpreted.

Returning to the approximate form of the Hamiltonian, we substitute equation (\ref{eqn:anharmonic}) for $r=1+\rho$ and expand the resulting expression as a power-series in $\Delta$. For the purposes of our analysis, it suffices to truncate this expansion at linear order in $\Delta$ for two reasons: first, retention of higher-order terms does not generate any slowly-varying harmonics and therefore does not contribute appreciably to the description of the system's principal resonance; second, the polynomial approximation for the piecewise-continuous potential employed in equation (\ref{eqn:Hpartan}) is only valid for small values of $\Delta$, meaning that a more precise representation of $(1/r^2)$ is unlikely to yield more accurate results.

\begin{figure*}[t]
\centering
\includegraphics[width=\textwidth]{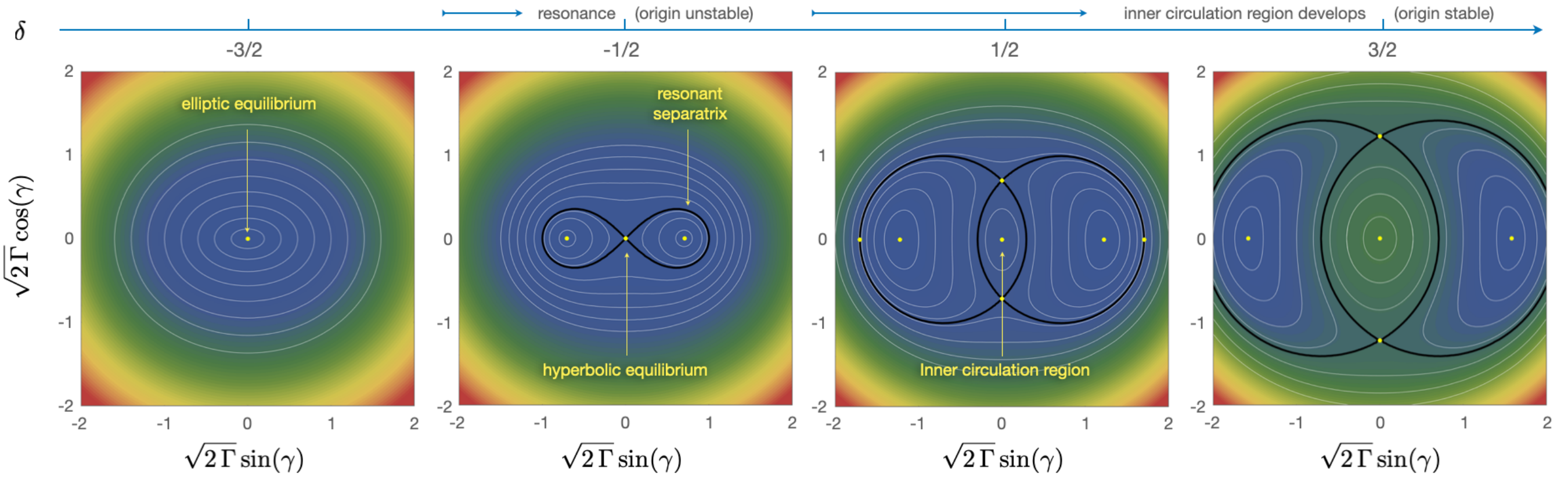}
\caption{Phase-space portraits of the Andoyer model, for a series of values of $\delta$. Equilibria of the Hamiltonian are shown with yellow dots, while the separatrix is shown with a black curve, where present. The topological character of the Andoyer Hamiltonian is fully determined by the resonance proximity parameter, $\delta$. Notably, in presence of a homoclinic curve, elliptic stability of the origin -- which translates to the existence of an inner circulation region of the phase-space -- necessitates that $\delta >0$.}
\label{F:Andoyer}
\end{figure*}

With these approximations in place, the Hamiltonian takes the form:
\begin{align}
\Hp&\approx\frac{\pu^2}{2}+\bigg(\frac{1}{1+b}\bigg)\frac{u^2}{2} + \bigg(\frac{b}{1+b}\bigg)\frac{u^4}{16}\nonumber \\
&-\bigg(\frac{b\,\Delta}{1+b}\bigg)\,u^2\,\cos(\omega\,t).
\label{eqn:Hpartucos}
\end{align}
To proceed further, we introduce action-angle coordinates $(\psi,\Psi)$, implicitly defined in terms of $(u,\pu)$:
\begin{align}
&u=\sqrt{2\,\Psi} \, \sin(\psi) & \pu=\sqrt{2\,\Psi} \, \cos(\psi).
\label{eqn:psiPsi}
\end{align}
To be precise, this change of variables simply amounts to a transformation from cartesian to canonical polar coordinates, and stems from a type-1 generating function of the form $\mathcal{F}_1=-(1/2)\,u^2\,\tan^{-1}(\psi)$. After some rearrangement, equation (\ref{eqn:Hpartucos}) becomes:
\begin{align}
\Hp&\approx\Psi-\bigg(\frac{b}{1+b}\bigg)\,\Psi\,\sin^2(\psi)+\bigg(\frac{b}{4\,(1+b)}\bigg)\,\Psi^2\,\sin^4(\psi)\nonumber\\
&-\frac{b\,\Delta}{1+b} \bigg(\Psi\,\cos(\omega\,t)-\frac{1}{2}\Psi\,\cos(2\,\psi-\omega\,t)\nonumber \\ 
&-\frac{1}{2}\Psi\,\cos(2\,\psi+\omega\,t) \bigg).
\label{eqn:HpartPsi}
\end{align}

In order to identify the slowly-varying harmonic in equation (\ref{eqn:HpartPsi}), let us consider $\Hp$ in the limit of vanishing $b$ and $\Delta$. In this regime, it is trivial to see that $\dot{\psi}=\partial\Hp/\partial\Psi\rightarrow 1$. Simultaneously, equation (\ref{eqn:Hcorepower}) dictates that $\omega=2$ for $b=0$. This means that the harmonic $\cos(2\,\psi-\omega\,t)$ will be slowly varying, and the associated angle must correspond to the principal resonance of the system. With this notion in mind, let us convert to a new set of action-angle variables $(\phi,\Phi)$ that reflect this property of the dynamics.

To accomplish this, we extend the phase-space of the system, by adding a dummy action, $\mathcal{T}$, conjugate to time \cite{Morbybook}, such that the new Hamiltonian becomes:
\begin{align}
\tHp=\Hp+\mathcal{T}.
\label{eqn:tHp}
\end{align}
Next, we employ a canonical contact transformation of variables, stemming from the type-2 generating function: 
\begin{align}
\mathcal{F}_2= (\psi-\omega\,t/2)\,\Phi + (t)\,\Xi. 
\label{eqn:tHp}
\end{align}
The transformation equations then trivially yield:
\begin{align}
&\phi=\frac{\partial\,\mathcal{F}_2}{\partial\,\Phi} =\psi-\frac{\omega}{2}\,t &\Psi=\frac{\partial\,\mathcal{F}_2}{\partial\,\phi}=\Phi \nonumber \\
&\xi=\frac{\partial\,\mathcal{F}_2}{\partial\,\Xi} = t & \mathcal{T}=\frac{\partial\,\mathcal{F}_2}{\partial\,t}=\Xi-\frac{\omega}{2}\,\Phi.
\label{eqn:contact}
\end{align}
In preparation for canonical averaging, let us write the transformed Hamiltonian $\tHp$ as $\tHp=\bHp+\hHp$, where
\begin{align}
\bHp=\bigg(\frac{1}{2}\bigg(\frac{2+b}{1+b} \bigg)-\frac{\omega}{2} \bigg)\,\Phi+\frac{3\,b}{32\,(1+b)}\,\Phi^2+\Xi
\label{eqn:bHp}
\end{align}
represents a trivially integrable kernel of the dynamics, and
\begin{align}
\hHp&=b\,\Phi\,\big(4\,(4-\Phi)\cos(2\,\phi+\omega\,\xi)+\Phi\,\cos(4\,\phi+2\,\omega\,\xi) \nonumber \\
&+16\,\Delta\,(\cos(2\,(\phi+\omega\,\xi))-2\,\cos(\omega\,\xi) \nonumber \\
&+\cos(2\,\phi))\big)/(32\,(1+b))
\label{eqn:hHp}
\end{align}
is a harmonic perturbation \cite{LightL}.

The terms in $\hHp$ that exhibit an explicit dependence on $\xi$ can be removed through a near-identity canonical transformation of variables. Adopting the Lie-series approach to perturbation theory \cite{1966PASJ...18..287H,1969CeMec...1...12D} we take this transformation to emanate from a generating Hamiltonian, $\chi$ (in this case, the un-averaged and averaged variables are related to one another through $\Phi=\mathcal{S}_{\chi}^{t}\bar{\Phi}$, $\phi=\mathcal{S}_{\chi}^{t}\bar{\phi}$, where $\mathcal{S}_{\chi}^{t}$ is the Lie series operator \cite{Morbybook}). Accordingly, carrying out the averaging procedure to first order, the homologic equation reads:
\begin{align}
\hHp+\{\bHp,\chi\}=\frac{b\,\Delta}{2\,((1+b))}\,\Phi\,\cos(2\,\phi),
\label{eqn:homologic}
\end{align}
where $\{,\}$ denote the Poisson bracket. For the problem at hand, equation (\ref{eqn:homologic}) is satisfied by:
\begin{align}
\chi&=-\frac{b\, (\Phi -4) \, \Phi }{16+b \, (3 \, \Phi +8)} \sin (2 \, \phi + \omega \, \xi) \nonumber \\
& + \frac{b \,\Phi ^2}{8\, (16+b \,(3 \, \Phi +8))}\sin (4 \,\phi+2 \,\omega \,\xi  ) \nonumber \\
&+\frac{4 \, b \,\Delta \, \Phi }{b \, (3 \, \Phi +8 \, \omega +8)+8 \,(\omega +2)}\sin (2 (\phi + \omega \, \xi )) \nonumber \\
&-\frac{b \, \Delta \, \Phi }{(b+1) \, \omega }\sin ( \omega \, \xi).
\label{eqn:chi}
\end{align}
After this reduction, particle dynamics are governed by the integrable Hamiltonian 
\begin{align}
\barHp&=\bigg(\frac{1}{2}\bigg(\frac{2+b}{1+b} \bigg)-\frac{\omega}{2} \bigg)\,\Phi+\frac{3\,b}{32\,(1+b)}\,\Phi^2\nonumber \\
&+\frac{b\,\Delta}{2\,((1+b))}\,\Phi\,\cos(2\,\phi).
\label{eqn:Hpfinal}
\end{align}
where we have dropped the now-constant action $\Xi$, and the averaged nature of the coordinates ($\Phi,\phi$) is implied. This expression agrees with equation (18) of ref. \onlinecite{BV}, although our derivation is different. 

The functional form of $\chi$ illuminates an important feature of the resonant structure of the dynamics encapsulated by Hamiltonian (\ref{eqn:Hpartan}) -- namely that conditions for $\chi$ to become singular are unphysical. In other words, the system at hand appears to naturally circumvent the \textit{small divisor problem}, at least to leading order (see Ch. 2.3 of \citet{Morbybook}). Qualitatively, this implies that as long as Hamiltonian (\ref{eqn:Hpartan}) constitutes a suitable approximation to the dynamical evolution jointly facilitated by Hamiltonians (\ref{eqn:Hcore}) and (\ref{eqn:Hpart}), the process of reducing $\Hp$ to an integrable form through time-averaging is justified under all physically relevant circumstances. In turn, this means that the emergence of broad chaotic layer is not expected within the parameter range of interest. 

This interpretation is consistent with the numerical results in refs. \onlinecite{Jameson1993,W98} (see also Figure \ref{F:uni} below). Nevertheless, we remark that the averaging procedure -- when not truncated at leading order (as done in equation \ref{eqn:homologic}) -- generates additional, higher-order resonant harmonics through ancillary terms in the averaged Hamiltonian of the form $\{\{\bHp,\chi\},\chi\}$, $\{\{\{\bHp,\chi\},\chi\},\chi\}$, etc \cite{LightL,Morbybook}. If the strength of such terms increases with $b$, chaotic motion can be driven by the modulation of the principal resonance's separatrix by such high-order resonances. We hypothesize that this is the reason why \cite{W98} observe chaotic motion in numerical simulations with $b=100$ but not for lower values of the space-charge parameter. 

\subsubsection{Particle-Core Interaction as an Andoyer Model}  \label{sec:andoyer}

Beyond providing a purely analytical explanation for numerically computed results, averaged Hamiltonian (\ref{eqn:Hpfinal}) is reminiscent of a mathematical pendulum. Unlike a conventional pendulum, however, the harmonic term in expression (\ref{eqn:Hpfinal}) possesses the D'Almbert characteristic of being multiplied by the action. In fact, equation (\ref{eqn:Hpfinal}) belongs to a family of so-called Andoyer Hamiltonians \cite{2007ASSL..345.....F}, which play a central role in the perturbative treatment of a broad array of astrophysical phenomena, including the evolution of stellar and planetary obliquities \cite{1987CeMec..40..345H, 1993Natur.361..608L, 2019ApJ...876..119M, 2013ApJ...778..169B}, secular interactions \cite{2011ApJ...739...31L, 2015ApJ...799..120B}, as well as orbital resonances \cite{HL83,1976ARA&A..14..215P}.

In general, Andoyer Hamiltonians have the form \cite{1984CeMec..32..127B}:
\begin{align}
\mathcal{K}&=(2\,k-5)(1+k\,\delta)\,\Gamma+k\,\Gamma^2\nonumber \\
&+(3-k)^{1+k/2}\,\Gamma^{k/2}\,\cos(k\,\gamma),
\label{eqn:Andoyer}
\end{align}
where $k$ is an integer greater than zero, $(\gamma,\Gamma)$ are action-angle variables, and $\delta$ is the resonance proximity parameter -- a number that quantifies how close the system is to exact commensurability. Importantly, $k$ dictates the possible number of stable resonant libration regions in phase-space, and for the problem at hand, $k=2$. Moreover, the topological character of the phase-space portrait of $\mathcal{K}$ is regulated exclusively by $\delta$, and it is straightforward to show that depending on the value of $\delta$, the qualitative picture lies in one of three regimes: (i) for $\delta<-1$, $\mathcal{K}$ only possesses a single elliptic equilibrium point at the origin, and the dynamics are essentially equivalent to that of a harmonic oscillator; (ii) for $-1\leqslant\delta<0$, the equilibrium point at $\Gamma=0$ is rendered hyperbolic, and a figure-8 resonant separatrix emerges; (iii) for $\delta\geqslant0$, an inner circulation region develops, and the separatrix takes on the shape of vesica piscis. To illustrate this, we show the phase-space portraits of Hamiltonian (\ref{eqn:Andoyer}) for $\delta =-3/2,-1/2,1/2,3/2$ in Figure (\ref{F:Andoyer}).

Let us rewrite the particle-core interaction model in the form of equation (\ref{eqn:Andoyer}), with an eye towards expressing $\delta$ in terms of $b$ and $\Delta$. The procedure of converting equation (\ref{eqn:Hpfinal}) into expression (\ref{eqn:Andoyer}) consists fo two consecutive steps: scaling of the actions, followed by a change of unit of time. First, we take $\Gamma=\Phi/\eta$ and $\mathcal{K}'=\barHp/\eta$. Upon direct substitution, we require the coefficient upfront the $\Gamma^2$ term to differ from the harmoinc term by a factor of $2$, as dictated by equation (\ref{eqn:Andoyer}) for $k=2$. This immediately yields an equation for $\eta$:
\begin{align}
\frac{3\,b\,\eta}{32\,(1+b)}=\frac{b\,\Delta}{1+b},
\label{eqn:eta}
\end{align}
which reduces to $\eta=32\,\Delta/3$.

\begin{figure}[t]
\centering
\includegraphics[width=\columnwidth]{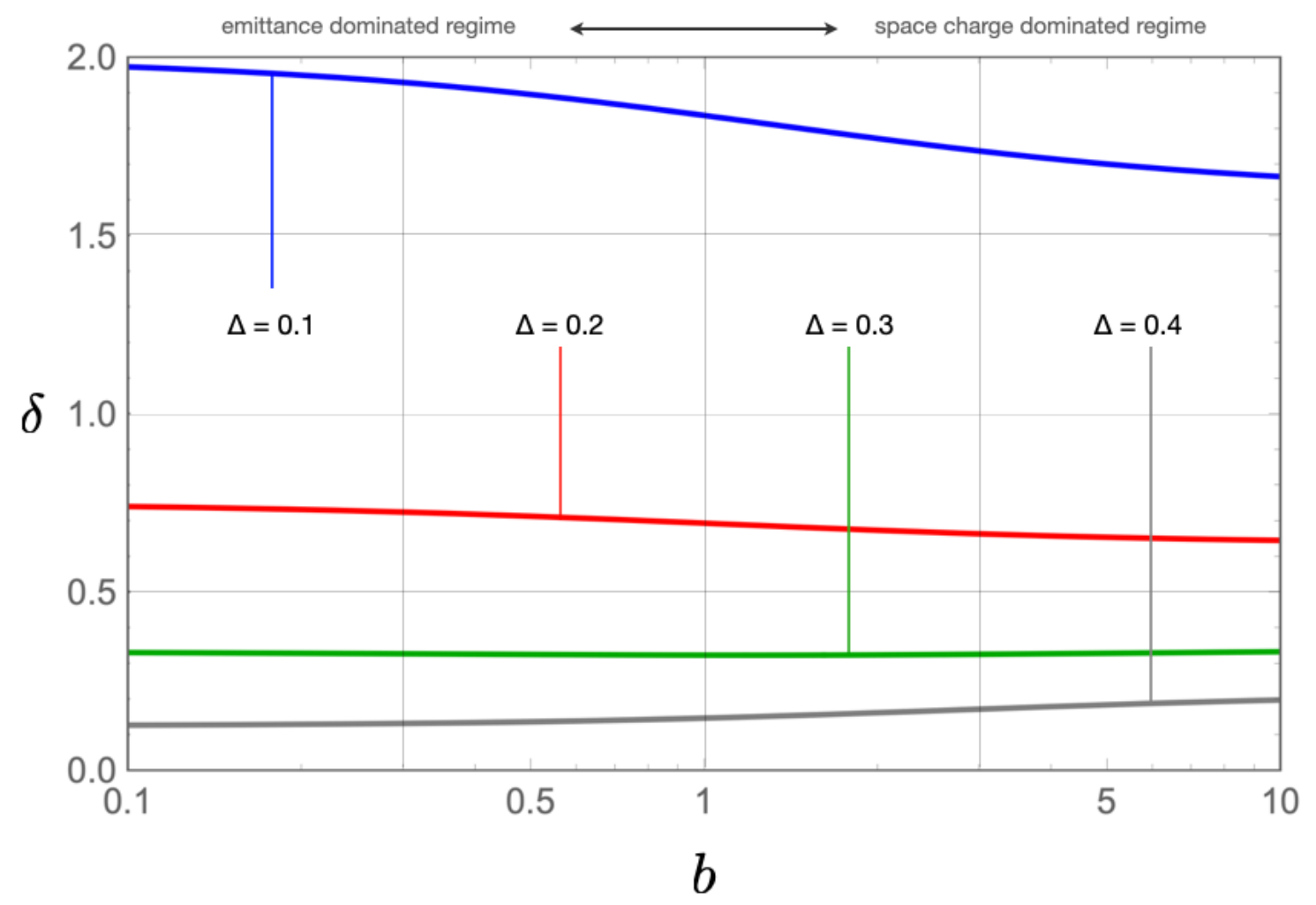}
\caption{Resonance proximity parameter, $\delta$, as a function of the space charge metric, $b$. The four depicted curves correspond to core oscillation amplitudes of $\Delta = 0.1$ (Blue), $\Delta = 0.2$ (Red), $\Delta = 0.3$ (Green), and $\Delta=0.4$ (Gray). Within the parameter range of interest, $\delta$ is always positive, implying that qualitatively, the phase-space portrait of the uniform particle-core interaction model will resemble the two RHS panels of Figure (\ref{F:Andoyer}).}
\label{F:delta}
\end{figure} 

\begin{figure*}[tbp]
\centering
\includegraphics[width=\textwidth]{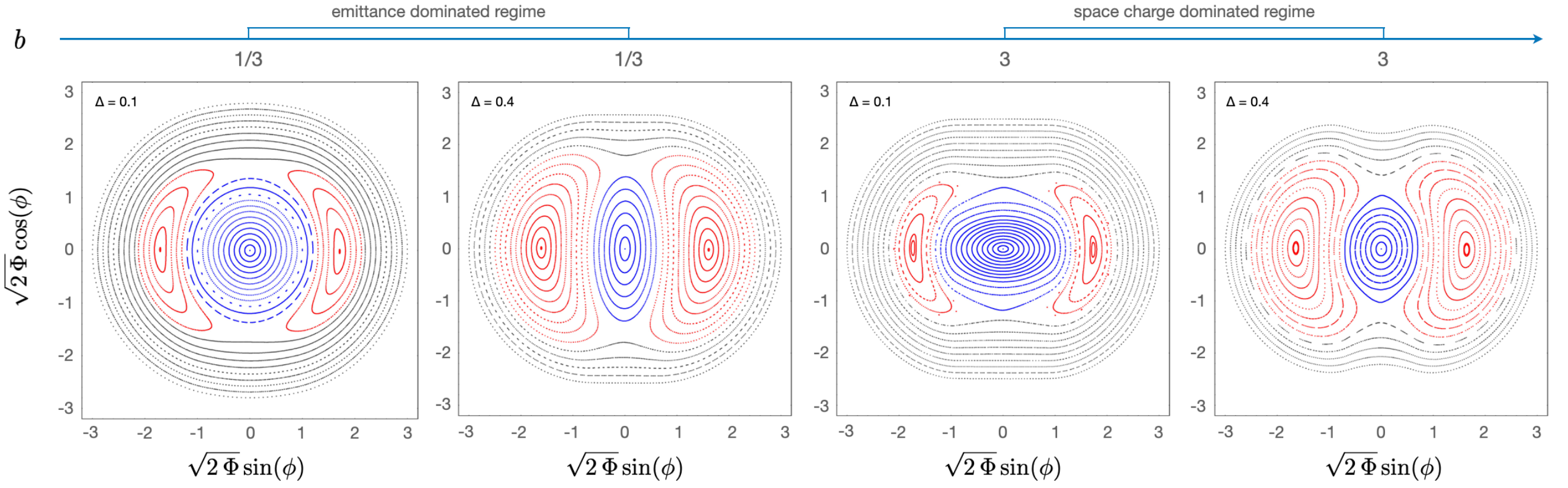}
\caption{Stroboscopic surfaces of section of particle motion, with a uniformly charged core. The two LHS panels of the Figure have $b=1/3$ and lie in the emittance dominated regime, while the panels on the RHS have $b=3$ and reside in the space charge dominated regime. Trajectories that occupy the inner and outer circulation regions of the phase-space diagram are shown in blue and gray respectively. Resonant orbits are shown in red. In all cases, the qualitative character of the surface of section is identical to that of the Andoyer model with $\Delta>0$. Moreover, even for large core pulsation amplitude ($\Delta=0.4$), particle motion remains quasi-periodic, as indicated by the fact that all trajectories fall on invariant tori.}
\label{F:uni}
\end{figure*} 

Finally, we set $\mathcal{K}=\mathcal{K}'\,(2(1+b))/(b\,\Delta)$, which reduces the Hamiltonian to the form given by equation (\ref{eqn:Andoyer}), with the resonance proximity parameter given by:
\begin{align}
\delta&=\frac{b^2 \left(\Delta ^2-\Delta +\omega -1\right)+b\, (3\, \omega -2 \,\Delta -4)+2 \,(\omega -2)}{2\, b\, (b+2) \, \Delta } \nonumber \\
&\approx\frac{(\omega-2)+b\,(\omega-1)}{2\,b\,\Delta}-\frac{1}{2}+\mathcal{O}(\Delta).
\label{eqn:delta}
\end{align}
It is illuminating to examine the limiting regimes of equation (\ref{eqn:delta}). In the space-charge dominated case of $b\gg1$, the resonance proximity parameter $\delta\approx(\sqrt{2}-1+\Delta^2-\Delta)/(2\,\Delta)$ is positive definite. In the emittance dominated regime of $b\ll1$, $\delta\approx(1-2\,\Delta)/(4\,\Delta)$, meaning that the existence of a stable inner circulation region of phase-space occupied by the core -- granted by the condition that $\delta>0$ -- is also guaranteed as long as $\Delta<1/2$. 

Although the possibility that $\delta$ can become negative for sufficiently large values of the beam pulsation amplitude -- causing the $\Phi=0$ fixed to become hyperbolically unstable -- is intriguing, we remind the reader that our entire analysis is predicated on the assumption that $\Delta$ is small. To this end, simultaneous numerical integration of equations of motion stemming from Hamiltonians (\ref{eqn:Hcore}) and (\ref{eqn:Hpart}) demonstrates that for low values of $b$, the $(u,\pu)=(0,0)$ initial condition remains stable, even for large values of $\Delta$. This suggests that the potential for $\delta$ to become smaller than zero insinuated by equation (\ref{eqn:delta}) is an artifact of the $\Delta\ll1$ assumption.

Figure (\ref{F:delta}) shows the resonance proximity parameter for $\Delta = 0.1 - 0.4$ as a function of $b$. Notably within our adopted range of beam parameters, $\delta$ is bracketed by $0\leqslant\delta\leqslant2$. Importantly, the fact that $\delta$ is of order unity despite large variations in underlying physical parameters of the beam, signifies a pronounced insensitivity of the phase-space portrait to the assumed values of the core pulsation amplitude, $\Delta$, and the beam space-charge parameter, $b$. In other words, equation (\ref{eqn:delta}) suggests that under all physically plausible conditions, the phase-space portrait of the particle-core interaction model will always resemble the two right-hand-side panels of Figure (\ref{F:Andoyer}). To further illustrate this point, in Figure (\ref{F:uni}) we show a series of stroboscopic surfaces of section, generated from direct integration of equations of motion stemming from Hamiltonian (\ref{eqn:Hpart}). Indeed, the qualitative agreement between the integrable approximation and numerical experiments is satisfactory.

\section{A Gaussian Beam} \label{sec:gaussianbeam}

Although the results presented in the proceeding section may illuminate the underlying structure of the phase-space portrait of a particle embedded within a uniformly charged beam, they do not yield any qualitative physical insights that were not already evident within the established understanding of the standard particle-core interaction model. That is, the notion of the existence of resonant separatrixes that can transport particles from the outer edges of the beam core to greater radial separations was already evident in the pioneering work of \citet{G94} and was further demonstrated from numerical grounds by \citet{L94b} and \citet{W98}. In fact, the only genuinely novel result of the analytic calculations presented in the previous section is an explanation for why the basic aspects of the particle's phase-space portrait are persistently insensitive to the assumed values of beam parameters. 

Actual beams, however, are not uniformly charged, as assumed in the preceding section. Rather, a broad collection of experimental measurements (see ref. \onlinecite{2016arXiv160802456P} and the references therein) indicate that within real high-intensity linacs, beam cores take on a profile that is closely approximated by a Gaussian distribution. One example of such an observation of a Gaussian profile of a space-charge-dominated beam stems from the SNS linac \cite{Plum12}, where tune depression achieves the value of $0.6$ \cite{2001pac..conf.2902B}, corresponding to space charge parameter $b \approx 1.8$. Another example is the observation of a starkly non-uniform beam profile in the J-PARC linac \cite{2016JKPS...69.1005M, Miyao}, where tune depression reaches the value of $0.5$ \cite{Liu19} corresponding to space charge parameter $b = 3$. 


It is important to understand that in real machines, the shape of the beam profile is determined by a multitude of competing effects. Specifically, space-charge forces themselves act to flatten the beam distribution \cite{1993PhRvL..71.2911R}, while other phenomena, including misalignments of accelerating and focusing elements, transverse-longitudinal coupling in the RF field, irregularities and instabilities of RF and focusing fields, as well as excitation of higher-order modes in the accelerating structures, act to widen the distribution. Cumulatively, experimental results indicate that these phenomena result in diffusive, Gaussian-type profile. Correspondingly, a Gaussian variant of the particle-core interaction model represents a more self-consistent description of particle dynamics.

As we show below, the more realistic case of a Gaussian beam exhibits greater qualitative complexity. Particularly, within the context of a Gaussian beam, the onset of large-scale stochastic motion can occur for physically plausible parameters, and the resulting phase-space mixing can even connect the radial excursions along the separatrixes to a larger section of the beam core. Deriving specific conditions for the appearance of chaos and the delineation of the associated irregular dynamics is the primary goal of this section. However, in order to set the stage for our analysis of chaotic dynamics, it is instructive to first derive an integrable model for a Gaussian beam, along the same lines as what was done above. 

\subsection{A Semi-Analytic Model} \label{sec:semianmodel}

\begin{figure}[b]
\centering
\includegraphics[width=\columnwidth]{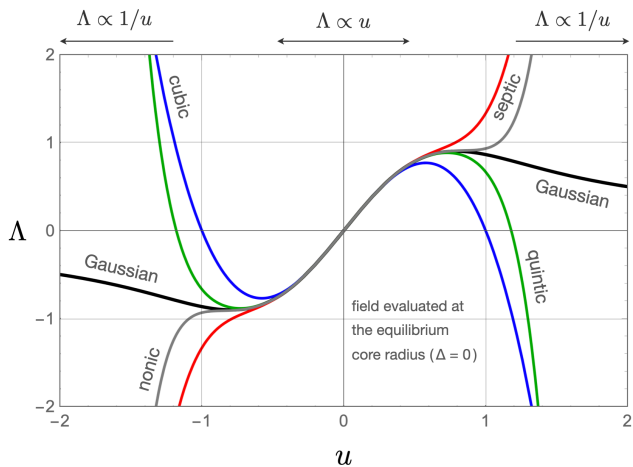}
\caption{The field of a Gaussian beam, $\Lambda$, evaluated setting the core radius to its equilibrium value of $r=1$, is shown as a black curve. Polynomial approximations to $\Lambda$ of various degrees are shown with multi-colored curves. Clearly, power-series expansions fail to capture the form of $\Lambda$ for $u\gtrsim1$, even at a crude level.}
\label{F:polyGauss}
\end{figure}

The electromagnetic field of a beam with a Gaussian charge distribution has the form:
\begin{align}
\Lambda=\frac{1}{u}\,\bigg(\frac{b}{1+b} \bigg)\,\bigg(1-\exp\bigg(-2\,\frac{u^2}{r^2}\bigg) \bigg).
\label{eqn:Gfield}
\end{align}
We assume that $r$ still obeys equation (\ref{eqn:anharmonic}) but is now interpreted as the 2-rms radius of the core. In doing so, we are assuming that beam oscillations are dominated by space-charge forces, and are ignoring all auxiliary effects that contribute to shaping the Gaussian profile of the beam. Nevertheless, this interpretation is consistent with experimental data reported by \citet{2016arXiv160802456P}, where the beam radius is shown to pulsate, while the beam profile remains consistently Gaussian.  Unlike the field of a uniform beam, the function (\ref{eqn:Gfield}) smoothly varies from $\Lambda\propto u$ for $u\ll r$ to $\Lambda\propto 1/u$ for $u\gg r$. Nevertheless, a power-series expansion $\Lambda\approx(b/(1+b))(2\,u/r^2-2\,u^3/r^4+...)$ does not yield an adequate approximation for equation (\ref{eqn:Gfield}) for $u\gtrsim r$, even if the degree of the polynomial is high (Figure \ref{F:polyGauss}). Consequently, for the sake of quantitative accuracy, here we proceed with the averaging method maintaining the expression for $\Lambda$ in closed form.

\begin{figure*}[t]
\centering
\includegraphics[width=\textwidth]{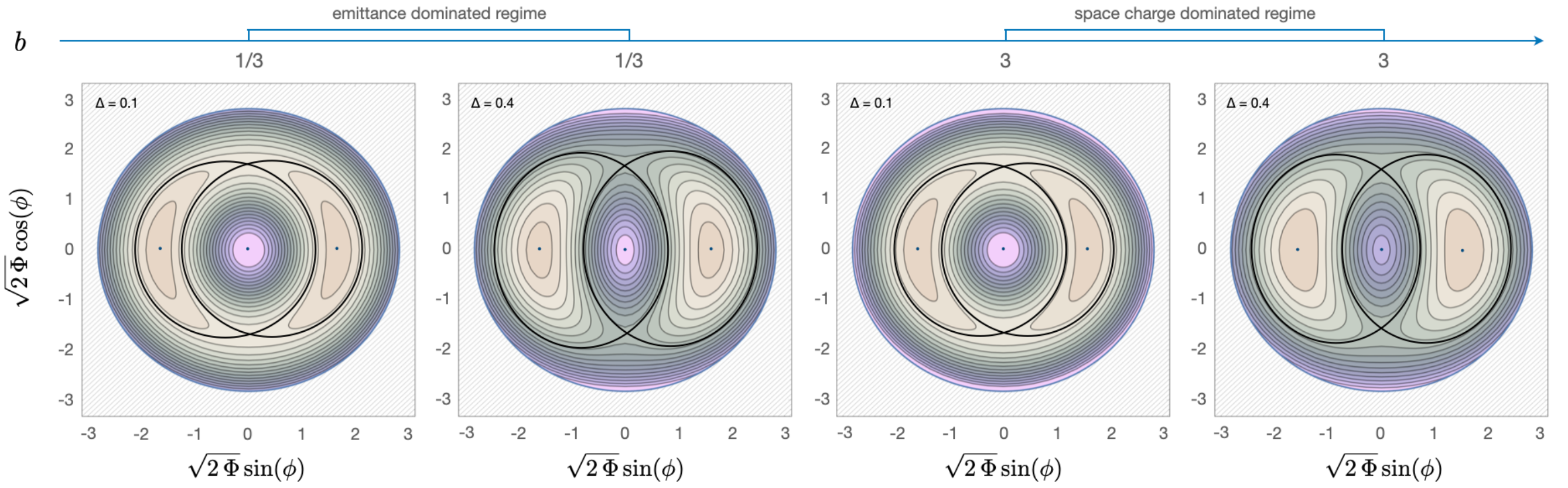}
\caption{Numerically averaged phase-space portraits of $\barHp$, depicting the resonance in $(\Phi,\phi)$, given a Gaussian charge distribution of the core. As in Figure (\ref{F:uni}), the two LHS and RHS panels lie in the emittance dominated and space charge dominated regimes, respectively. Qualitatively, the outlines of this resonance are indistinguishable from the Andoyer model with $\delta>0$.}
\label{F:num_ave}
\end{figure*} 

The Hamiltonian corresponding to particle motion within a Gaussian beam is:
\begin{align}
\Hp&=\frac{\pu^2}{2}+\frac{u^2}{2}+\mathcal{T} \nonumber \\
&+\frac{b}{1+b}\bigg(\frac{\Ei(-2\,u^2/r^2)}{2} - \ln(u) \bigg),
\label{eqn:Hgauss}
\end{align}
where $\mathcal{T}$ is a dummy action conjugate to time and $\Ei$ is the exponential integral. Changing variables in accord with canonical transformations outlined in equations (\ref{eqn:psiPsi}) and (\ref{eqn:contact}), the above expression becomes:
\begin{align}
\Hp&=\bigg(1-\frac{\omega}{2}\bigg)\,\Phi+\Xi-\frac{b\,\ln\big(\sqrt{2\Phi}\sin(\phi+\omega\,\xi/2) \big)}{1+b} \nonumber \\
&+\frac{b}{2\,(1+b)}\,\Ei\bigg(-\frac{4\,\Phi\,\sin^2(\phi+\omega\,\xi/2)}{r^2} \bigg),
\end{align}
and we remind the reader that $\xi=t$.

Rather than attempting to reduce $\Hp$ to a one-degree-of-freedom Hamiltonian through the Lie series approach (which entails solving the homologic equation \ref{eqn:homologic} as in section \ref{sec:intapp}), here we instead take advantage of separation of timescale inherent to the ($\Phi,\phi$) and ($\Xi,\xi$) degrees of freedom. In particular, because $\dot{\phi}\sim 1-\omega/2\ll\dot{\xi}=1$, the evolution of ($\Phi,\phi$) is \textit{adiabatic} with respect to radial pulsation of the core \cite{1982CeMec..27....3H,1984PriMM..48..197N}. In this regime, we are justified in directly averaging the Hamiltonian over a single oscillation cycle in $\xi$:
\begin{align}
\barHp&=\frac{\omega}{2\,\pi}\oint \big(\Hp - \Xi \big)\,d\xi.
\label{eqn:directave}
\end{align}

Because this procedure reduces $\barHp$ to an integrable system semi-analytically, in order to illuminate the associated dynamics, we can simply project the level curves of $\barHp$ onto the ($\Phi,\phi$) plane. Figure (\ref{F:num_ave}) shows a sequence of phase-space portraits of $\barHp$ for $b=1/3$ and $b=3$ (we have also computed analogous Figures for $b=0.1,1,10$, and found them to be very similar to those depicted in Figure \ref{F:num_ave}). Crudely speaking, this exercise suggests that the qualitative character of averaged dynamics within a Gaussian and a uniform beam are identical: in both cases, an inner circulation region is encircled by a vesica piscis-shaped separatrix, with two resonant equilibria residing on the $u$-axis. Moreover, the pronounced insensitivity of the phase-space portraits to the beam parameters $b$ and $\Delta$ appears to be in line with the functional form for the resonance proximity parameter suggested by equation (\ref{eqn:delta}).

A more cursory inspection of the Hamiltonian, however, reveals that even at a basic level, the picture cannot be so simple, since a prominent additional resonance -- previously removed from $\Hp$ via the averaging process (\ref{eqn:directave}) -- emerges as an important driver of particle dynamics for sufficiently large values of $b$. To demonstrate this, let us return to equation (\ref{eqn:Hgauss}), set $\Delta=0$ (such that $r=1$, which renders $\Hp$ autonomous), and examine its equilibrium. In particular, it is instructive to consider the Hessian matrix of $\Hp$, evaluated at the $(u,\pu)=(0,0)$ fixed point. Since there is no $u-\pu$ coupling in the Hamiltonian, $\partial^2\Hp/\partial u\,\partial\pu = 0$. It is further trivial to see that $\partial^2\Hp/\partial\pu^2 = 1$. The determinant of the Hessian matrix is then simply
\begin{align}
\mathcal{D}&=\frac{\partial^2\Hp}{\partial u^2} = \lim_{u\rightarrow 0}\bigg[1+\frac{b}{1+b}\frac{e^{-2\,u^2}(e^{2\,u^2}-4\,u^2-1))}{u^2} \bigg] \nonumber\\
&=1-\frac{2\,b}{1+b}<0 \ \ \ \ \ \mathrm{if} \  b>1.
\label{eqn:determinant}
\end{align}

\begin{figure*}[t]
\centering
\includegraphics[width=\textwidth]{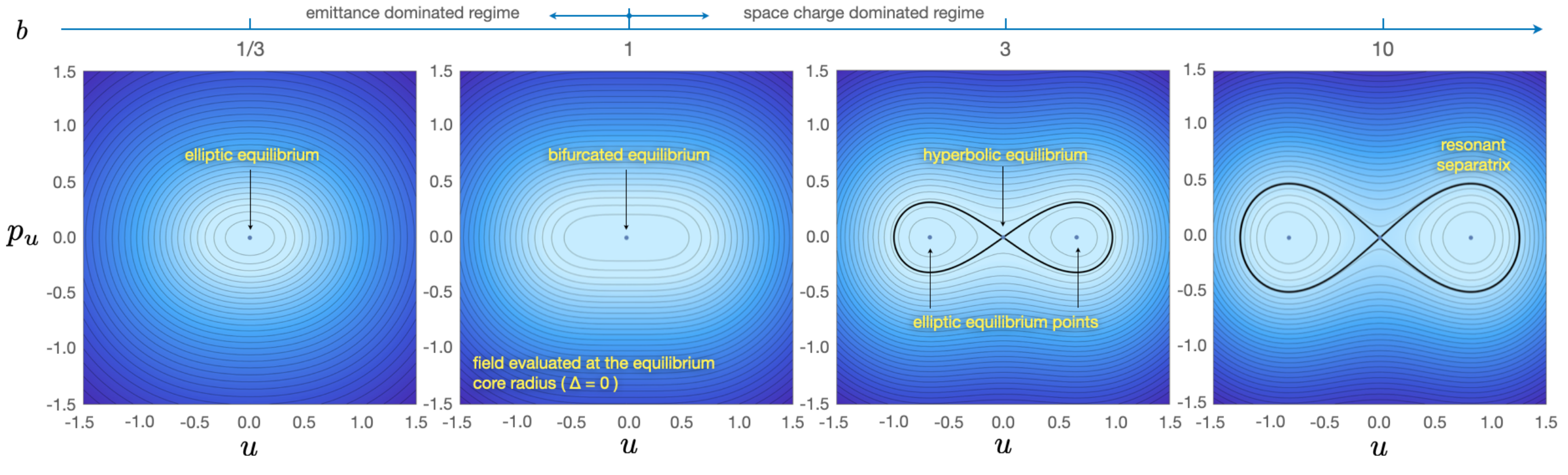}
\caption{Phase-space portraits of $u=\sqrt{2\,\Psi}\,\sin(\psi),\pu=\sqrt{2\,\Psi}\,\cos(\psi)$ dynamics within a Gaussian beam for $\Delta=0$. A nonlinear resonance appears for $b>1$, rendering the origin of the phase-space diagram hyperbolically unstable. Importantly, this resonance is distinct from that generated by the breathing mode of the core (when $\Delta>0$), and does not emerge in the uniformly charged variant of the particle-core interaction model. The phase-space topology of this resonance is equivalent to that of the Andoyer model with $-1<\delta<0$.}
\label{F:psires}
\end{figure*}

This rudimentary analysis shows that as $b$ crosses the threshold of unity (which coincides with the transition from the emittance dominated regime to the space-charge dominated regime), the origin of the $(u,\pu)$ phase-space transitions form being an elliptic fixed point (corresponding to a minimum of $\Hp$) to a hyperbolic fixed point (corresponding to a saddle point of $\Hp$), signaling the appearance of a resonant separatrix for $b>1$. Importantly, these dynamics arise completely independently of the breathing mode of the core. The emergence of a figure-eight separatrix, along with new elliptic equilibria at $|u|>0$, $\pu=0$ is shown in Figure (\ref{F:psires}), where we plot the phase-space portrait of $\Hp$ for $b=1/3,1,3,10$, setting $\Delta=0$. An important consequence of the appearance of this resonance is that when $\Delta$ is made sufficiently large, this resonance can overlap the resonance shown in Figure (\ref{F:num_ave}), driving prominent chaotic motion within the core.

We note that herein lies an important distinction between the Gaussian and uniformly-charged variants of the particle-core interaction model. That is, contrary to the dynamics encapsulated by the Hamiltonian (\ref{eqn:Hgauss}), linear stability of the origin in a $\Delta=0$ uniformly charged beam is trivially ascertainable. In particular, from equation (\ref{eqn:Hpart}), it is easy to see that near the $(u,\pu)=(0,0)$ point, the Hamiltonian has the form of a harmonic oscillator: $\Hp=\pu^2/2+(1/(1+b))\,u^2/2$, implying a determinant of the Hessian matrix that is positive definite. Consequently, a nonlinear resonance of the kind depicted in Figure (\ref{F:psires}) can never arise within the uniformly-charged model, meaning that chaotic motion within the core is only expected to arise within the context of a Gaussian beam.

\begin{figure*}[t]
\centering
\includegraphics[width=\textwidth]{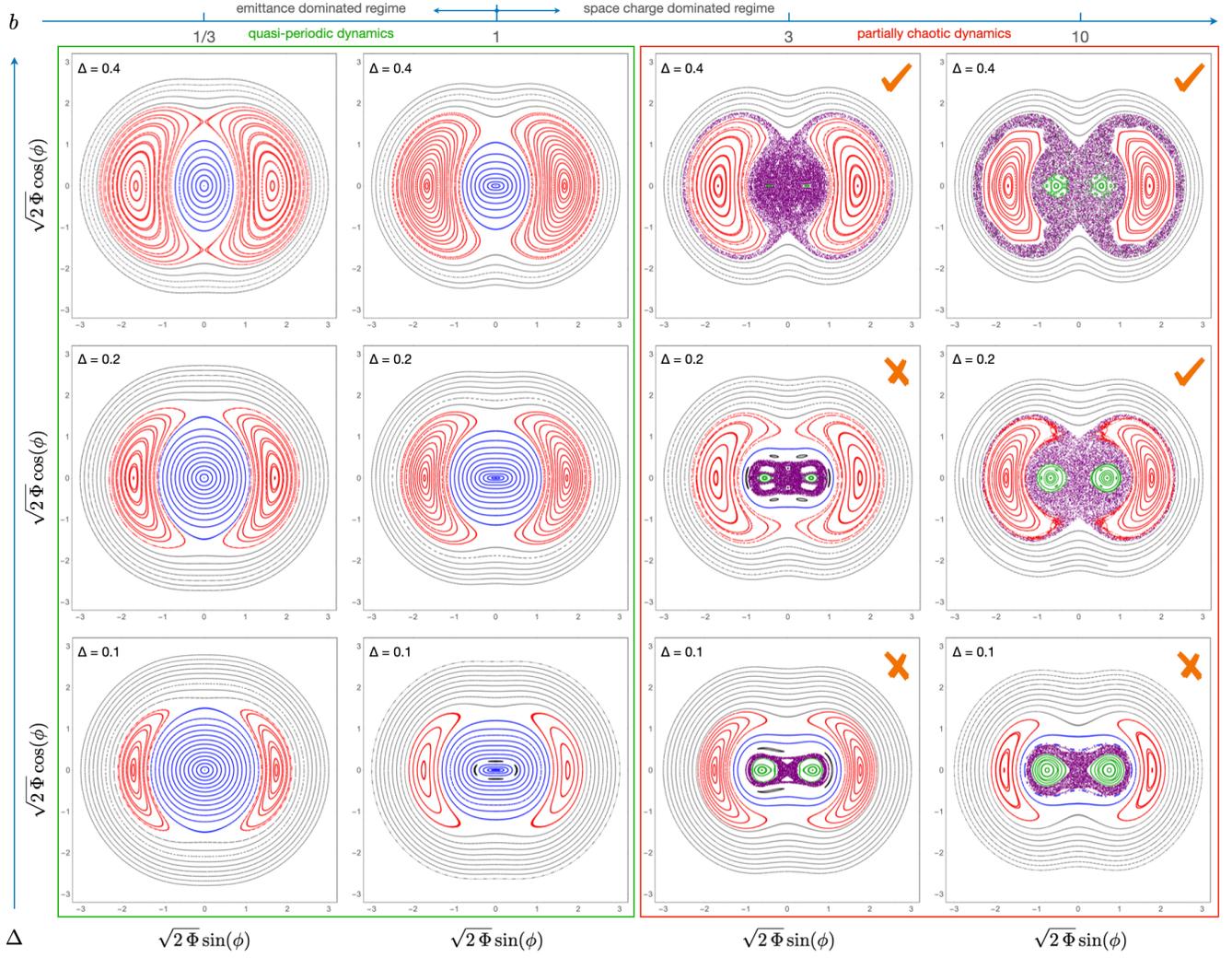}
\caption{A gallery of stroboscopic surfaces of section of particle dynamics within a Gaussian beam. The same color scheme as that employed in Figure (\ref{F:uni}) is also adopted here, with the added element of chaotic trajectories being depicted in purple. In the emittance dominated ($b=1/3$) and transitionary ($b=1$) regimes, dynamical evolution within a Gaussian beam is largely equivalent to that of the uniformly charged beam, and follows the well-known narrative of the particle-core interaction model. On the other hand, in the space charge-dominated regime (corresponding to panels with $b=3$ and $b=10$), a discernible chaotic layer emerges close to the origin, facilitating unpredictable particle evolution. Importantly, for sufficiently large core pulsation amplitudes, separatrixes of the ``intrinsic" Gaussian resonance (in $\psi$) and the ``breathing" resonance (in $\phi$) overlap strongly, connecting the core and the halo through rapid stochastic transport. RHS panels where chaotic core-halo mixing ensues are labeled with a checkmark, whereas those where the chaotic layer is confined to the core are marked with a diagnoal cross.}
\label{F:Gausssurf}
\end{figure*}

Numerically computed surfaces of section shown in Figure (\ref{F:Gausssurf}) largely confirm this expectation. In particular, unlike the case of a uniformly-charged beam, where the trajectories that comprise the core remain largely quasi-periodic, here a conspicuous stochastic layer appears close the origin for $b\gtrsim1$. These stroboscopic plots can be compared to equivalent surfaces of section of the uniformly charged beam depicted in Figure (\ref{F:uni}). Importantly, in both cases, the surfaces of section were obtained through direct integration of the equations of motion derived from Hamiltonians (\ref{eqn:Hcore}, \ref{eqn:Hpart}, \ref{eqn:Hgauss}), and are not influenced by any of the approximations (e.g., polynomial fit to the field, etc) invoked above. 

A crucial consequence of the appearance of a stochastic layer within the beam core is that it allows for chaotic transport to ensue over a broader phase-space domain. That is, for sufficiently large values of $\Delta$, this layer connects to the principal resonance's separatrix, effectively linking the beam core and the halo through chaotic mixing. Let us now quantify the onset of chaos within a Gaussian particle-core interaction model, and derive the relevant criteria from analytic grounds.

\subsection{Onset of Chaos} \label{sec:onset}

A quantitative understanding for conditions under which dynamical systems begin to exhibit stochastic behavior dates back to the pioneering work of \citet{1960JPlPh...1..253C, Chirikov1979}, and by now, the essential statement of the Chirikov criterion is widely known: in order for global chaos to ensue, the sum of the half-widths of a pair of resonances in action space must exceed the distance between the resonances' equilibria. One approach towards calculating this criterion in the context of charged particle motion within a Gaussian beam, is to compute the locations and widths of individual resonant separatrixes semi-analytically, as shown in Figures (\ref{F:num_ave}) and (\ref{F:psires}). A considerably less accurate (see Figure \ref{F:polyGauss}), but more qualitatively informative approach is to adopt a series-expansion of the governing Hamiltonian, with an eye towards reducing $\Hp$ to the simplest possible model that still captures irregular dynamics, and estimating the Chirikov criterion in a purely analytical manner. This is the strategy we adopt in this section.


As a starting point, let us write down an approximate variant of Hamiltonian (\ref{eqn:Hgauss}), truncating the polynomial expansion of $\Lambda$ at cubic order:
\begin{align}
\Hp&=\frac{\pu^2}{2}+\frac{u^2}{2}-\frac{b}{1+b}\bigg(\frac{u^2}{r^2}-\frac{u^4}{2\,r^4} \bigg).
\label{eqn:Hgaussexp}
\end{align}
This Hamiltonian is clearly reminiscent of equation (\ref{eqn:Hpartan}), although the quartic term in the above expression is notably stronger, and is dependent on $r$. We now follow a procedure similar to that outlined in section \ref{sec:intapp}. We substitute equation (\ref{eqn:anharmonic}) for $r$, but to keep the model as compact as possible, we expand the resulting expression to leading order in $\Delta$. Moreover, in spirit of a series-expansion approach, upon transforming to action-angle variables ($\Psi,\psi$) via equation (\ref{eqn:psiPsi}), we neglect all harmonics proportional to $\Psi^2$, retaining only those that are linear in $\Psi$. This reduces the Hamiltonian to the following simplified form:
\begin{align}
\Hp&\approx\frac{\Psi}{1+b}+\frac{3\,b\,\Psi^2}{4\,(1+b)}\nonumber \\
&+\frac{b\,\Psi}{1+b}\big(1+2\,\Delta\,\cos(\omega\,t)\big)\cos(2\,\psi).
\label{eqn:Hmodpend}
\end{align}

Importantly, Hamiltonian (\ref{eqn:Hmodpend}) has a similar mathematical structure to the Hamiltonian of a \textit{modulated pendulum}. As a mechanical system, a modulated pendulum is nothing more than a mass suspended via a rigid rod, whose base hangs on a spring that oscillates with frequency $\omega$. In turn, oscillations of the base result in a periodic variation in the effective acceleration due to gravity \cite{LightL}. Although qualitatively simple, depending on system parameters and initial conditions, the modulated pendulum can display a broad range of behavior, including strictly periodic motion, quasi-periodic resonant trajectories, motion that lies on KAM tori, as well as full-fledged deterministic chaos. For this reason, the modulated pendulum is routinely adopted as a paradigm for stochastic dynamics  \cite{1989PhyD...40..265B,1991PhyD...54..135H,1991Nonli...4..615E, Morbybook}.

The application of the Chirikov resonance overlap criterion to the modulated pendulum is straightforward. If we momentarily ignore the dependence of the cosine term in $\Hp$ upon $\Psi$, it is easy to see how sinusoidal variation its strength would give rise to a total of three resonances. In particular, because $(1+2\,\Delta\,\cos(\omega\,t))\,\cos(2\,\psi)=\cos(2\,\psi)+\Delta\,\cos(2(\psi-\omega\,t/2))+\Delta\,\cos(2(\psi+\omega\,t/2))$, the fixed points of these resonances would be separated in action space (equivalently, frequency space) by $\omega/2$. Accordingly, global chaos would ensue when the half-widths of these resonant multiplets become comparable to $\omega/2$ such that the separatrixes cross.

All of the dynamical features of a conventional modulated pendulum also naturally arise within the context of Hamiltonian (\ref{eqn:Hmodpend}). The analogy, however, is inexact in one important respect: while a standard modulated pendulum always encompasses three resonant harmonics with well-defined equilibrium conditions, the system at hand possesses the D'Almbert characteristic, which limits the physically meaningful domain of the action $\Psi$ to positive values (such that $u$ and $\pu$ are real). Additionally, the D'Almbert characteristic modifies the distance between the resonant equilibrium points. This means that in order for chaos to ensue, not only do the resonances have to overlap, but at least two of the corresponding equilibria have to reside at $\Psi>0$. As we already demonstrated in the previous sub-section via equation (\ref{eqn:determinant}), this condition translates to a requirement that $b>1$. We remind the reader that this constraint can be written as $\sigma/\sigma_0<1/\sqrt{2}$, corresponding to the transition from emittance-dominated regime to the space-charge dominated regime. Assuming that this constraint is satisfied, let us now write down the approximate criterion for the onset of chaos.


Unlike the width of the unperturbed resonance in $\psi$ (which only exists for $b>1$ but is largely independent of $\Delta$), the width of the resonance corresponding to $\phi=\psi-\omega/2$ depends on $\Delta$. Consequently, a reasonable way to formulate the Chirikov criterion is to equate the distance between the stable equilibrium points of the $\psi$ and $\phi$ resonances to the half-width of the resonance in $\phi$. Let us now compute the relevant quantities. The equilibrium action of the unperturbed $(\Psi,\psi)$ dynamics, $[\Psi]$, follows from Hamilton's equations in the $\Delta\rightarrow0$ limit:
\begin{align}
&\frac{d\,\psi}{dt}=\frac{\partial\,\Hp}{\partial\,\Psi}=\frac{1}{1+b}+\frac{3\,b\,[\Psi]}{2\,(1+b)}+\frac{b\,\cos(2\,[\psi])}{1+b}=0 \nonumber \\
&\frac{d\,\Psi}{dt}=-\frac{\partial\,\Hp}{\partial\,\psi}=-\frac{2\,b\,[\Psi]\,\sin(2\,[\psi])}{1+b}=0.
\label{eqn:equilibriumeqns}
\end{align}
From inspection, it is evident that the equilibrium value of the angle lies at $[\psi]=\pi/2,3\pi/2$ such that $\cos(2[\psi])=-1$. The value of the unperturbed equilibrium action is then
\begin{align}
[\Psi]=\frac{2\,(b-1)}{3\,b}.
\label{eqn:Psieq}
\end{align}
The functional form of $[\Psi]$ is yet another reminder of the fact that chaotic dynamics can only appear for $b\geqslant1$: for values of $b$ smaller than unity, resonance overlap is precluded by $[\Psi]$ taking on a negative value, such that that the resonant equilibrium resides on the imaginary $(u,\pu)$ plane.

The value of the equilibrium action $[\Phi]$ can be computed in a similar manner. Upon transforming variables according to equations (\ref{eqn:contact}) and dropping (i.e., ``averaging away") $\xi$-dependent terms, Hamilton's equations give:
\begin{align}
&\frac{d\,\phi}{dt}=\frac{\partial\,\Hp}{\partial\,\Phi}=\frac{1-\Delta\,b}{1+b}-\frac{\omega}{2}+\frac{3\,b\,[\Phi]}{2\,(1+b)}=0,
\label{eqn:equilibriumeqns2}
\end{align}
where we have set $\cos(2\,[\phi])=-1$ by analogy with equation (\ref{eqn:equilibriumeqns}). After mild rearrangement, the distance between resonant equilibria in action-space reads:
\begin{align}
[\Phi]-[\Psi]=\frac{\omega+b\,(\omega+2\,\Delta-2)}{3\,b}.
\label{eqn:distance}
\end{align}

The remaining step is to compute the resonance half-width $\delta\Phi$. In the pendulum approximation -- which is justifiable for the $\phi$-resonance as long as $\delta\Phi\lesssim [\Phi]$ \cite{1983Icar...56...51W} -- we can substitute $[\Phi]$ for the value of the action upfront the cosine term in $\Hp$. This gives \cite{Morbybook}:
\begin{align}
\delta\Phi\approx 2\sqrt{\bigg(\frac{b\,\Delta\,[\Phi]}{1+b} \bigg) \bigg(\frac{2\,(1+b)}{3\,b}\bigg)}=\sqrt{\frac{8\,\Delta\,[\Phi]}{3}}.
\label{eqn:halfwidth}
\end{align}
Recalling that the existence of the resonance in the ($\psi,\Psi$) degree of freedom requires that $b>1$, the Chirikov criterion $([\Phi]-[\Psi])/\delta\Phi\lesssim1$ can now be written as 
\begin{align}
\frac{\omega+b\,(\omega+2\,\Delta-2)}{2\,\sqrt{2}\sqrt{b\,\Delta\,(\omega+b\,(\omega+2\,\Delta)-2)}} \lesssim 1 \ \ \mathrm{and} \ \ b>1.
\label{eqn:Chirikov}
\end{align}
Although it is possible to express $b$ as a function of $\Delta$ (or alternatively $\Delta$ as a function of $b$) in closed form, the resulting expressions are cumbersome, and do not yield any additional insight that is not already evident in the above expression. 

Figure (\ref{F:Lyapunov}) depicts the Chirikov criterion (\ref{eqn:Chirikov}) as a thick cyan curve on the $(\Delta,b)$ plane. Upon inspection, two physically important characteristics of this threshold are immediately evident. First, for diminishing values of $\Delta$, progressively higher values of $b$ are necessary to drive chaotic motion. This relationship stems from the $\sim1/\sqrt{\Delta}$ dependence in equation (\ref{eqn:Chirikov}), and simply insinuates that in the limit of vanishing $\Delta$, $\Hp$ reduces to an integrable form, implying perfectly regular dynamics. 

Second, for $\Delta>(\sqrt{3}-1)/3\approx0.24$, the expression (\ref{eqn:Chirikov}) suggests that a stochastic layer can emerge for a value of $b$ below unity. This is, however, an artifact of our pendulum-like formulation of the Chirikov criterion, and as we have already discussed above, $b>1$ constitutes a necessary but insufficient condition for resonance overlap. Stated differently, our analysis suggests that for core pulsation amplitude in excess of $\sim25\%$, particle dynamics within a Guassian beam become chaotic essentially as soon as the space-charge dominated regime is reached.

Before leaving this sub-section, let us briefly comment on the suitability of the assumed profile of the beam core. Aforementioned experimental support for a Gaussian beam profile aside, the emergence of a prominent nonlinear resonance within a stationary beam -- and bonafide stochasticity within a pulsating one -- begs the question of self-consistency. That is, if initially stationary particles positioned in the center of the beam begin to execute significant radial excursions as soon as the beam becomes space charge-dominated, surely the particle-core interaction model implies that a Gaussian distribution cannot be maintained indefinitely. In fact, this is precisely what is observed in idealized multi-particle numerical simulations: on rather short timescales, an initially Gaussian beam exhibits the propensity to self-regulate, rapidly approaching a uniform charge distribution \cite{1994tdcp.book.....R}. It is important to understand however, that this tendency is suppressed in real machines, underscoring the limitations of numerical experiments that only include space charge as a means of beam profile modulation, and fail to account for other important physical effects. In particular, lattice misalignments and beam mis-steering, together with nonlineaities arising from the RF field and focusing elements, act to restore the (approximately) Gaussian nature of the beam, suggesting that while chaotic diffusion within the particle core interaction model is driven by nonlinear space-charge forces, the very nonlinearity of these effects are maintained through extrinsic forcing.

\subsection{Numerical Experiments} \label{sec:numexp}
The analytic conclusions of the preceding discussion can be tested and extended with the aid of numerical simulations -- an exercise we now carry out. The series of questions we seek to answer in the remainder of this section are three-fold. First, we wish to quantify the validity of equation (\ref{eqn:Chirikov}) within the context of our system. Second, we aim to compute the characteristic Lyapunov exponent pertinent to charged particle dynamics. Finally, we wish to address the practically important issue of mapping out conditions under which chaotic diffusion within the core ceases to be bounded, and links to more extended radial excursions, facilitating an accelerated formation of a beam halo. To address these questions, we performed the following suite of numerical experiments.

On a grid of beam parameters spanning $b\in(0,10), \Delta\in(0.01,0.4)$, we computed $100$ stroboscopic surfaces of section, akin to those shown in Figure (\ref{F:Gausssurf}). To populate each surface of section, we chose 30 initial conditions lying on the $u-$axis, spanning the range $u_0\in(0,1-\Delta)$. We note that this choice is made strictly for definitiveness, and we have confirmed that our results do not exhibit any significant dependence on the distribution of initial conditions. Recalling that our surfaces of section correspond to a phase of core pulsation when $r$ is minimized, these initial conditions correspond to ions that originate inside the core with zero radial velocity. Every trajectory was evolved for 1000 core pulsation periods.

In conjunction with each integration of the equations of motion, we additionally evolved a second, shadow trajectory (initialized in close proximity to the main trajectory at $\acute{\pu}=\epsilon,\acute{u}=u_0+\epsilon$, where $\epsilon=10^{-6}$) through integration of linearized equations of motion (see appendix \ref{appendix:megno}). Keeping track of the distance between the primary and shadow trajectories in phase-space, $\nu=\sqrt{(\pu-\acute{\pu})^2+(u-\acute{u})^2}$, we compute the Lyapunov coefficient, $\lambda$, of the trajectory. To be more specific, we computed the so-called \textit{mean exponential growth factor of nearby orbits} \cite{2000A&AS..147..205C}:
\begin{align}
Y=\frac{2}{t}\int_{0}^{t}\frac{\dot{\nu}}{\nu}t'\,dt',
\label{eqn:MEGNO}
\end{align}
which known to converge more rapidly than the standard renormalization method for computing $\lambda$ \cite{1980Mecc...15....9B}. To this end, we note that the running time-average of $Y$ tends towards $\bar{Y}\sim2$ if the trajectory is regular, and towards $\bar{Y}\sim\lambda\,t/2$ if the trajectory is chaotic. 

\begin{figure}[t]
\centering
\includegraphics[width=\columnwidth]{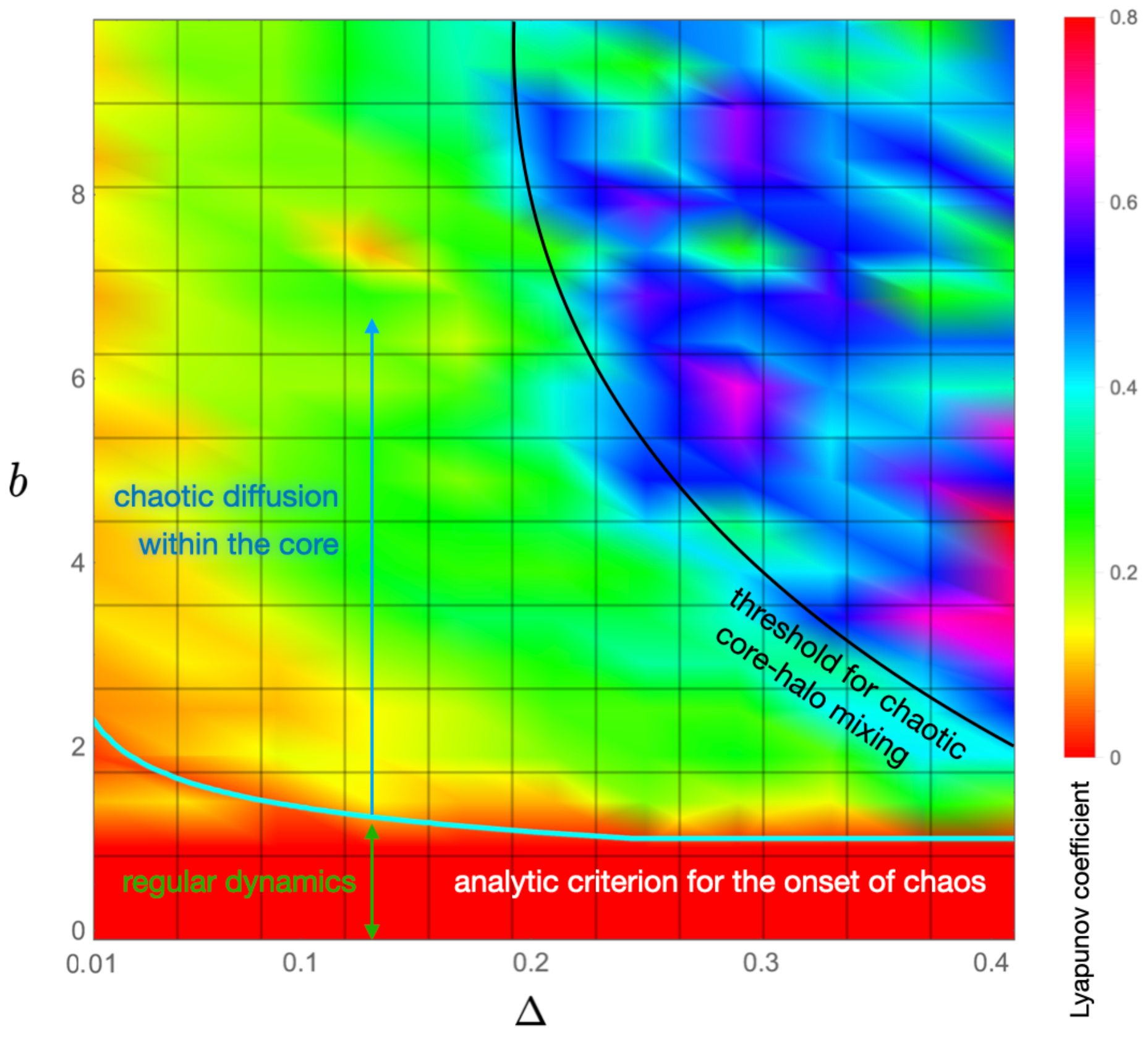}
\caption{Regular and chaotic dynamics within the Gaussian particle-core interaction model. The background contour plot informs the topological character of the stroboscopic surface of section by reporting the Lyapunov coefficient, $\lambda$, where chaotic motion is detected. The analytical Chirikov criterion for the onset of stochasticity (equation \ref{eqn:Chirikov}) is marked with a cyan curve. Note that while the Chirikov criterion yields $b\approx 2.3$ for the lowest numerically tested beam pulsation amplitude of $\Delta=0.01$, this curve turns sharply towards $b\rightarrow\infty$ as $\Delta\rightarrow0$, where the dynamics becomes integrable. Finally, the threshold for chaotic mixing between the core and the halo is outlined with a black curve.}
\label{F:Lyapunov}
\end{figure}

If all initial conditions on the surface of section yield quasi-period dynamics, we record $\lambda=0$ for the given combination of $\Delta$ and $b$. If, on the other hand, some trajectories result in chaotic motion, we record the mean value of $\lambda$, obtained by averaging the Lyapunov coefficients of chaotic initial conditions (i.e., initial conditions that do not give stochastic behavior are excluded). The resulting distribution of $\lambda$ is shown as a background contour-plot on Figure (\ref{F:Lyapunov}). Although imperfect, the analytic criterion for the onset of chaos (equation \ref{eqn:Chirikov}) clearly tracks the boundary between regular and stochastic dynamics. Thus, in spite of a series of relatively crude approximations that were made in reducing the Gaussian particle core interaction model to a simplified modulated pendulum-like Hamiltonian (\ref{eqn:Hmodpend}), the resulting criterion for the onset of chaos displays satisfactory agreement with numerical simulations. 

Beyond the binary question of whether or not a given combination of beam parameters yields chaotic motion or not, the value of $\lambda$ itself tells an intriguing story. That is, the distribution of $\lambda$ is clearly bimodal, with moderate values of $\Delta$ and $b$ corresponding to $\lambda\sim0.1-0.3$ and large values of $\Delta$ and $b$ corresponding to $\lambda\gtrsim0.5$. Crucially, the boundary between these two domains appears rather sharp, and corresponds to the onset of chaotic linkage of the separatrixes of $\phi$ and $\psi$ resonances. To check this, we kept track of whether or not trajectories that originate within the core attain a value of $u$ that is greater than that of the stable equilibrium point of the $\phi$-resonance (computed using semi-analytical averaging as in section \ref{sec:semianmodel}) on the surface of section, and derived an approximate threshold for such mixing. This threshold is shown on Figure (\ref{F:Lyapunov}) as a black curve, and crudely tracks the boundary between slow and rapid diffusion. Thus, the region of parameter space characterized by the larger value of the Lyapunov coefficient also corresponds to the regime where halo formation is facilitated by enhanced stochastic mixing of trajectories, which joins the beam core and halo through rapid chaotic transport, along the outlines of resonant separatrixes in phase-space.

\section{Summary} \label{sec:summary}
In this work, we have carried out a perturbative, as well as numerical investigation of the particle-core interaction model of halo formation, with an eye towards understanding the role played by large-scale chaotic diffusion, and have quantified the conditions for its onset. A key result of our analysis is the demonstration that the emergence of globally stochastic dynamics within the framework of the particle-core interaction paradigm hinges upon the assumed charge distribution within the core. That is, while particle motion within a uniform beam is largely regular, an appreciable fraction of phase-space occupied by the core can become chaotic given physically plausible parameters if the charge distribution of the beam is Gaussian.

The analytic calculations presented in sections \ref{sec:partdyn} and \ref{sec:semianmodel} explain this difference from qualitative grounds. In particular, for the case of a uniform beam, we show that the particle-core interaction model can be reduced to an integrable Andoyer-type Hamiltonian through canonical perturbation theory, and that the associated generating Hamiltonian is strictly non-singular, meaning that the averaging procedure is justified -- at least to first order in the resulting series \cite{Morbybook} -- for all physically plausible beam parameters. This elucidates why chaotic motion is confined to a thin band around the separatrix when the charge-distribution of the core is uniform \cite{W98,BV}.

The case of a Gaussian beam is considerably more complex. While a uniform beam-like qualitative picture holds in the emittance-dominated regime, our analysis reveals that as soon as the system crosses over into the space-charge dominated regime, a second principal resonance emerges in phase space in addition to the usual breathing mode of the particle-core interaction model. This resonance renders the center of the beam hyperbolically unstable. Modulation of the corresponding figure-$\infty$ separatrix by the core's breathing mode gives rise to chaotic dynamics within the core. Moreover, for sufficiently high values of the beam's space charge parameter $b$ and beam pulsation amplitude $\Delta$, the two principal resonances can overlap strongly, leading to enhanced chaotic mixing between the core and the halo (Figure \ref{F:Lyapunov}). Indeed, the boundaries of model parameter space within which deterministic stochasticity reigns are predictable, and are given by a simple analytic expression (\ref{eqn:Chirikov}). Accordingly, the analysis presented herein outlines a theoretical framework within the context of which more detailed numerical and experimental results can be interpreted.



\begin{acknowledgments}
K.B. is grateful to Caltech as well as the David and Lucile Packard Foundation for their generous support. Y.B. is grateful to the Los Alamos National Laboratory and the U.S. Department of Energy for their generous support, under US DOE contract 89233218CNA000001.
\end{acknowledgments}

\section*{Data Availability Statement}
The data that support the findings of this study are available from the corresponding author upon reasonable request.

\begin{appendix}

\section{An approximate solution for the pulsation of the core}  \label{appendix:anharmonic}
In the vicinity of the $(\rho,\prho)=(0,0)$ equilibrium, Hamilton's equations of motion -- applied to expression (\ref{eqn:Hcore}) -- give:
\begin{align}
&\frac{d\,\rho}{d\,t}=\frac{\partial\,\Hc}{\partial\,\prho}=\prho \nonumber \\
&\frac{d\,\prho}{d\,t}=-\frac{\partial\,\Hc}{\partial\,\rho}=-\omega^2\,\rho+\zeta^2\rho^2.
\label{eqn:HamcoreEOM}
\end{align}
Following standard practice \cite{LL,LightL}, we note that for low-amplitude oscillations the linear term in $\rho$ will dominate (yielding a harmonic oscillator), and consider the trial solution
\begin{align}
\rho\approx\Delta\,\cos(\omega\,t+\varphi)+\zeta^2\,f,
\label{eqn:trial}
\end{align}
where $f$ is a yet-to-be-determined function of time.

Writing equations (\ref{eqn:HamcoreEOM}) as a single second order differential equation for $\rho$, and substituting the trial solution, we have:
\begin{align}
&\zeta^2\,\frac{d^2\,f}{d\,t^2}-\Delta\,\omega^2\,\cos(\omega\,t+\varphi)=-\Delta\,\omega^2\,\cos(\omega\,t+\varphi) \nonumber \\
&+\Delta^2\,\zeta^2\,\cos^2(\omega\,t+\varphi) - \zeta^2\,\omega^2\,f +\zeta^6\,f^2 \nonumber\\
&+2\,\Delta\,\zeta^4\cos(\omega\,t+\varpi).
\end{align}
To second order in $\zeta$, the above equation of motion simplifies to:
\begin{align}
&\frac{d^2\,f}{d\,t^2}=\Delta^2\,\cos^2(\omega\,t+\varphi) - \omega^2\,f.
\end{align}
This expression is satisfied by the solution:
\begin{align}
f=\frac{\Delta^2\,\zeta^2}{2\,\omega^2}-\frac{\Delta^2\,\zeta^2}{6\,\omega^2}\,\cos(2\,(\omega\,t+\varphi)).
\end{align}
Equation (\ref{eqn:anharmonic}) is thus obtained. Although it is straightforward to extend this procedure to derive higher-order corrections to this solution, Figure (\ref{F:an}) indicates that for our purposes a second-order approximation is sufficient.

\section{Linearized Equations of Motion} \label{appendix:megno}
A well-known hallmark of chaotic evolution is the sensitive dependence on initial conditions, which manifests through exponential divergence of nearby trajectories. Since this divergence only emerges locally, it is standard practice to linearize the equations of motion and compute the phase-space distance between the primary and shadow trajectories $\delta u=\acute{u}-u$, $\delta\pu=\acute{\pu}-\pu$ according to:
\begin{align}
&\frac{d\, \delta u}{dt}=\frac{\partial^2\,\Hp}{\partial\pu^2}\,\delta\pu+\frac{\partial^2\,\Hp}{\partial u\,\partial \pu}\,\delta u \nonumber \\
&\frac{d\, \delta\pu}{dt}=-\frac{\partial^2\,\Hp}{\partial u\,\partial\pu}\,\delta\pu-\frac{\partial^2\,\Hp}{\partial u^2}\,\delta u 
\end{align}

For the problem at hand, these equations of motion take on a particularly simple form. As we already saw in section \ref{sec:semianmodel}, the mixed derivatives vanish because there is no $u-\pu$ coupling in the system. Moreover, the second derivative of $\Hp$ with respect to $\pu$ is just unity, and $\partial^2\Hp/\partial u^2$ is given by equation (\ref{eqn:determinant}). As a result, we have: 
\begin{align}
&\frac{d\,\delta\pu}{dt}=-\bigg(1+\frac{b}{1+b}\frac{e^{-2\,u^2}(e^{2\,u^2}-4\,u^2-1))}{u^2} \bigg)\delta u \nonumber \\
&\frac{d\,\delta u}{dt}=\delta\pu
\end{align}

\end{appendix}


\end{document}